# Rewriting results in the language of compatibility

Valentin Amrhein[1*] and Sander Greenland[2]

[1]Department of Environmental Sciences, Zoology, University of Basel, Basel, Switzerland
[2]Department of Epidemiology and Department of Statistics, University of California, Los Angeles, CA, USA
*v.amrhein@unibas.ch, @vamrhein

18 February 2022

Muff *et al.* [1] suggest that "when each paper only contributes a piece of evidence in the cumulative process of creating knowledge", practical decision makers should act on this cumulative knowledge rather than on single studies. But sadly, cumulative knowledge has a huge gap: 'Statistically non-significant' results, arising from a point estimate near the null or a wide interval estimate, are underrepresented in the published literature [2]. Muff *et al.*'s [1] proposal to rewrite results sections in the language of evidence may help to bridge this gap by allowing authors to interpret their *P*-values as graded measures of evidence about a certain finding or effect, so that also results with larger *P*-values could be described as providing 'weak' or 'little' evidence rather than as being merely 'statistically non-significant'. We agree that this would be a step forward and that a call to describe evidence provided by *P*-values across a range of magnitudes could help bring more of the larger *P*-values into the published literature.

**More good than harm**
Unfortunately, 'evidence' is a loaded and disputed term. We are concerned that a description such as "the data did not have any evidence about the direction of any association", as in Muff *et al.*'s [1] Table 1, is too easily misunderstood as "there was no association" [3]. Also, statements about "strong evidence *for*" an association, as in their Tables 1 and 2, are objectionable because *P*-values do not measure support but only indicate the amount of information *against* a tested (null) hypothesis or model [4].

Nonetheless, we think that, in practice, the proposed graded "evidence language" [1] would do more good than harm. One reason is that, in regular cases, *P*-values correlate well with 'evidence' as measured by likelihood ratios (e.g., we can obtain an approximate *P*-value by doubling the maximum likelihood ratio and looking that up in a $\chi^2$ table [5, section 9.3]. Another reason is that the 'evidence' scale and wording may be vague enough to largely avoid the overconfidence and categorization associated with the language of 'significance', 'confidence', 'credibility', or 'error control'. Those and similar terms suggest a mathematical rigor that often inadequately captures the uncertainties of real data generation, particularly in observational studies.

Muff *et al.* [1] also note that confidence intervals should not be used for binary decisions based on single studies, for example because random noise alone can cause them to vary dramatically from sample to sample even when all assumptions are correct [6]. In empirical research, assumptions will almost always be violated to some degree; hence, a 95% confidence interval cannot realistically be claimed to have a 95% coverage of the 'true effect' [7]. An interval estimate should instead be used for giving a sense of random variability (noise) in the point estimate rather than for categorical statements about the position of the true effect or about the probability that intervals would capture it.

**Compatibility and estimation rather than evidence**

To this end, a more careful and appropriate language would describe magnitudes of possible true effects as being more or less 'compatible' with our data, given our statistical model.

For example, a large *P*-value indicates there is little information against the tested (null) hypothesis, so in this sense it signals high compatibility of the hypothesis with the data under the assumptions used for the test. Under those assumptions, a hypothesis with a larger *P*-value is thus more compatible with the data than are other hypotheses with smaller *P*-values derived using the same data and testing method. A traditional 95% confidence interval can then be interpreted as a 'compatibility interval' [7,8] summarizing many test results, because the interval includes all hypotheses with $p > 0.05$ given the data [5, section 7.2]. Such intervals therefore summarize the possible effect sizes that are most compatible with our data according to their *P*-values, given our assumptions.

In this way, the language of compatibility shifts the focus away from a statement about just one hypothesis to a statement across a range of hypotheses, thus aiding depiction of uncertainty. While formulations containing 'evidence' are already customary in the literature, claims that the concept is captured by one or another statistical measure are quite contested [5, Chapter 2]. Compared with 'evidence' language, 'compatibility' language allows us to retain traditional and precisely defined methods while encouraging more accurate thinking about what the outputs mean.

**Whatever language you choose, be open and modest**

Because single studies contribute only incrementally to cumulative knowledge, not only is it "irrelevant whether an individual study was 'significant' or not" [1], but most any inference about an effect from a single study alone will be irrelevant in the face of all evidence. To aid summarizing knowledge in future meta-analyses, it is more helpful to report our studies by following phrases such as "we here fully report all analyses and all results"; "these are some of the possible biases in our study"; and "our data and code are openly available". In short, we should be thoughtful, open, and modest [9].

...